\documentclass[12pt]{iopart}
\usepackage{iopams,amsfonts,verbatim}

\usepackage{citesort}
\usepackage{graphicx}

\usepackage{tikz}
\usepackage{pgfplots}
\usepackage{makecell}
\usetikzlibrary{arrows}
\usetikzlibrary{shapes.geometric}
\usetikzlibrary{positioning}
\usetikzlibrary{calc}
\usetikzlibrary{decorations.pathmorphing}
\usetikzlibrary{fit}
\pgfdeclarelayer{background}
\pgfdeclarelayer{foreground}
\pgfsetlayers{background,main,foreground}
\usetikzlibrary{automata}
\usetikzlibrary{optikz}
\usetikzlibrary{decorations.markings}
\usetikzlibrary{patterns}
\usetikzlibrary{snakes}

   \newdimen \xsep \xsep =3cm %half distance between mirrors
    \newdimen \ysep \ysep =1.8cm %half distance between top and bottom levels

\begin{document}

\title{Minimization of Ion Micromotion in a Linear Paul Trap with a High Finesse Cavity}

\author{Boon Leng Chuah, Nicholas C. Lewty, Radu Cazan and Murray D. Barrett}

\address{Centre for Quantum Technologies and Department of
  Physics, National University of Singapore, 3 Science Drive 2, 117543 Singapore.}

\ead{cqtcbl@nus.edu.sg}

\date{\today}

\begin{abstract}
We demonstrate minimization of ion micromotion in a linear Paul trap with the use of a high finesse cavity. The excess ion micromotion projected along the optical cavity axis or along the laser propagation direction manifests itself as sideband peaks around the carrier in the ion-cavity emission spectrum. By minimizing the sideband height in the emission spectrum, we are able to reduce the micromotion amplitude to approximately the spread of the ground state wave function. This method is useful for cavity QED experiments as it allows for efficient 3-D micromotion compensation despite optical access limitations imposed by the cavity mirrors. We also show that sub-nanometer micromotion compensation is possible with our current system.
\end{abstract}


\maketitle

\section{Introduction}

Trapped ions have become an increasingly important technology for a wide range of applications including precision metrology \cite{rosenband,chou} and quantum information processing (QIP) \cite{wineland,home,wineland2,duan}. Due to the levels of precision demanded in these applications, it is important that the internal and motional degrees of freedom are well controlled. For instance, the quantum gate proposed by Cirac and Zoller requires the ion to be in motional ground-state for high fidelity operation \cite{cirac,schmidt}. 

In an ideal radio frequency (RF) trap, a cold ion is fixed at the zero of the RF electric field and no excess motion should be present. However, in practice, the presence of stray DC fields or a phase difference between the RF electrodes can induce excess micromotion. In the frame of the ion, this micromotion is equivalent to a modulation of the cooling laser and leads to sideband generation in the emission spectrum. Adverse effects include trap heating, reduction in the laser Rabi rate, and imperfect Raman-type state transfer \cite{wineland,jauregui,plata,berkeland,barrett}. Second order Doppler shifts due to excess micromotion are also a significant limitation to the attainable accuracy of atomic clocks. Thus, the detection and compensation of excess micromotion is an important requirement for many applications.

A variety of techniques for the minimization of micromotion have been discussed in the literature \cite{berkeland,raab,walther,ibaraki,narayanan}. While these techniques are widely used in ion trap experiments, their implementation can be hindered by limited optical access or cannot readily quantify the degree to which the micromotion is compensated. In addition, fluorescence techniques often require the RF drive frequency to be much larger than the linewidth of the optical transition used. This is not always easy to satisfy, particularly for heavier ions.

In this article, we present a method to minimize excess ion micromotion by using a high finesse cavity as a spectrum analyzer for light scattered into the cavity from a probe beam.  The moderate single atom cooperativity of the cavity enhances the amount of light scattered into the cavity and, in the presence of excess micromotion, frequency sidebands at the RF drive frequency in the cavity emission spectrum appear \cite{stute}. The heights of the sideband peaks allow us to directly measure the amplitude of the micromotion along two orthogonal directions and micromotion compensation is achieved by minimizing the sidebands. Our approach is applicable as long as the RF drive frequency ($\Omega$) is much greater than the cavity linewidth ($\kappa$), a condition easily fulfilled for most cavity QED experiments implemented with high finesse cavities.

\begin{figure}
\centering
\includegraphics{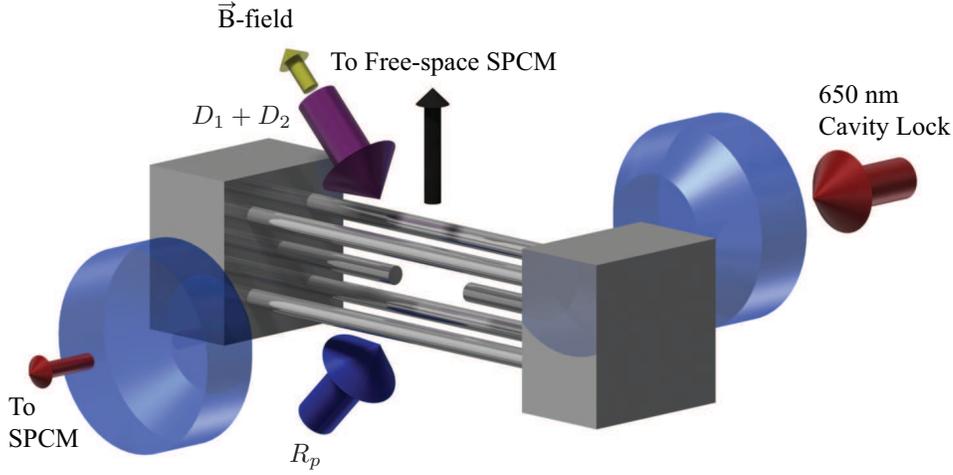}
\caption{\label{iso} The schematic of the setup. A single $^{138}\mathrm{Ba}^{+}$ is trapped at the RF trap center and coupled to a high finesse cavity. Two laser beams are used for fluorescence detection and Doppler cooling at $493\, \mathrm{nm}$ ($D_1$) and $650\, \mathrm{nm}$ ($D_2$) respectively, indicated by the purple arrow. A magnetic field is applied to define the quantization axis, indicated by the green arrow ($\hat{\mathbf{z}}$). The ion-cavity emission spectrum is probed by a $493\, \mathrm{nm}$ beam ($R_p$), indicated by the blue arrow ($\hat{\mathbf{x}}$). The photons emitted from the cavity are collected into a fiber-coupled single-photon counting module (SPCM). A CCD camera, interchangeable with another free-space SPCM, detects the fluorescence of the ion in the direction indicated by the black arrow. The cavity length is stabilized to a $650 \, \mathrm{nm}$ laser, indicated by a red arrow which is aligned to the cavity axis ($\hat{\mathbf{y}}$). }
\end{figure}

\section{The Model} \label{sec2}
We consider a set up in which an intra-cavity ion is probed transversely to the cavity as depicted in \Fref{iso}.  When the detuning, $\Delta$, of the probe from the atomic resonance is large relative to the linewidth, we can adiabatically eliminate the excited state.  This results in an effective Hamiltonian which, in the interaction picture, is given by
\begin{equation}
\label{hamiltonian}
H_I = \Omega_R \exp(-i\Delta_c t) a^{\dagger}+\Omega_R^{\ast} \exp(i\Delta_c t) a \, ,
\end{equation}
where $a$ is the cavity annihilation operator, and $\Delta_c$ is the probe detuning relative to the dispersively shifted cavity resonance.  The effective driving strength $\Omega_R$ determines the position dependent scattering of the probe into the cavity and is given by
\begin{equation}
\Omega_R=\frac{g \Omega_L}{\Delta} \exp(i k x)\sin\left(k y+\phi\right)
\end{equation}
where $g$ is the maximum ion-cavity coupling strength, $\Omega_L$ is the atom-probe coupling strength, $k$ is the wavenumber of the probe field, and $\phi$ determines the position of the ion along the cavity axis.  Without loss of generality, we can take the equilibrium position of the ion to be at $x=0=y$ and we consider micromotion $x(t)=x_m \cos\left(\Omega t\right)$ and $y(t)=y_m \cos\left(\Omega t\right)$ along the $x$ and $y$ directions respectively, where $\Omega$ is the RF drive frequency.  Expanding $\Omega_R$ to first order in $\beta_x=k x_m$ and $\beta_y=k y_m$ then gives the effective Hamiltonian
\begin{eqnarray}
\label{hamiltonian2}
\fl H_I = \frac{g \Omega_L}{\Delta}\exp(i\Delta_c t)\Bigg[\sin(\phi)\left(1+\frac{i}{2}\beta_x\left(\exp(i\Omega t)+\exp(-i\Omega t)\right)\right) \nonumber\\
+ \frac{1}{2}\beta_y\cos{\phi}\left(\exp(i\Omega t)+\exp(-i\Omega t)\right)\Bigg]a +\mathrm{h.c.}
\end{eqnarray}
Thus, the cavity emission will include sidebands at the RF drive frequency, $\Omega$. Provided $\Omega \gg \kappa$, where $\kappa$ is the field decay rate of the cavity, the rate of photons emitted from the cavity will be given by
\begin{equation}
I_c = I_{c0}\Bigg[\frac{\kappa^2}{\kappa^2+\Delta_c^2} + \frac{\beta_x^2}{4}\left(\frac{\kappa^2}{\kappa^2+(\Delta_c+\Omega)^2}+\frac{\kappa^2}{\kappa^2+(\Delta_c-\Omega)^2}\right)\Bigg]
\end{equation}
or
\begin{equation}
I_c=I_{c0}\frac{\beta_y^2}{4}\left(\frac{\kappa^2}{\kappa^2+(\Delta_c+\Omega)^2}+\frac{\kappa^2}{\kappa^2+(\Delta_c-\Omega)^2}\right)
\end{equation}
when the ion is located at the antinode ($\phi=\pi/2$) or node ($\phi=0$) respectively. In these equations, $I_{c0}$ is the rate of photons detected at the cavity output when the cavity is tuned to be resonant with the probe and $\phi=\pi/2$. Thus, $I_c(\Omega)/I_{c0}$, for each configuration, gives a direct measure of the micromotion amplitudes along the probe ($\hat{\mathbf{x}}$) and the cavity axis ($\hat{\mathbf{y}}$) directions.

To determine the limits of this approach to micromotion compensation, we first consider the case in which $\phi=\pi/2$. In this case the number of photons, $N_s$, collected at the micromotion sideband in an integration time $\tau$ is $N_s=\beta_x^2 I_{c0}\tau/4=\beta_x^2 N_c/4$ where $N_c$ is the number of photons collected at resonance.  Background counts at the RF sideband come from both off resonant scattering into the cavity and dark counts from the counting module.  If the micromotion features are well resolved, the background will be approximately constant around the sideband. Taking the mean counts of the background to be $N_b$ and assuming Poissonian statistics, the signal to noise ratio, $S$, will be then given by
\begin{equation}
S=\frac{\beta_x^2}{4} \frac{N_c}{\sqrt{N_b}} \, .
\end{equation}
Taking $S=1$ as the condition for minimum detectable $\beta_x$ gives
\begin{equation}
\label{betaxmin}
\beta_{x,\mathrm{min}}=2\sqrt{\frac{\sqrt{N_b}}{N_c}} \, .
\end{equation}
A similar expression holds for the minimum $\beta_y$ obtained when the ion is located at a cavity node, provided one uses the same value for $N_c$ as in \Eref{betaxmin}.

Typically $\Omega\gg\kappa$ such that the background counts would be dominated by dark counts from the counting module.  In this case the micromotion compensation improves with $\sqrt{N_c}$. Thus the degree of micromotion compensation will depend on the single atom cooperativity of the cavity, the free space scattering rate of the probe beam, and the level of cooling of the ion; all of which impact on the amount of probe light scattered into the cavity.  In addition we note that the degree of micromotion compensation has only a weak dependence on the integration time with $\beta_\mathrm{min}\sim \tau^{-1/4}$.

\section{The Experiment} \label{sec3}

\begin{figure}
\centering
\begin{tikzpicture}[scale=1,
      level/.style={line width=1.5pt,line cap=round},
     virtual/.style={line width=1.5pt,dashed,line cap=round},
      arrow/.style={very thick,line width=1pt,->,shorten <=0pt,shorten >=4pt},
      photon/.style={decorate, decoration={pre length=0.3cm,
                         post length=0.25cm, snake,amplitude=1mm,
                         segment length=3mm}, thick,line width=1pt,->,shorten <=4pt,shorten >=4pt},
     cavity/.style={draw, shape=vertical cavity, line width=1pt, optical element width=2*\xsep, optical element height=4cm},
     dimension/.style={very thick,line width=1pt,->,shorten <=0pt,shorten >=1pt},
     ]
\large

    \newdimen \levlength \levlength=2.5cm
    \newdimen \levlengthd \levlengthd=1.5cm

%\node[cavity, scale=1] at (0,0) (cav1) {};
\draw[level] (-\levlength/2,-\ysep) -- ++(\levlength,0) node[midway] (botmid2) {}  node[midway,yshift=0cm,xshift=-1.7cm] {$\mathrm{S}_{1/2}$};
\draw[level] (-\levlength/2,\ysep) -- ++(\levlength,0) node[midway,yshift=0cm,xshift=-1.7cm] {$\mathrm{P}_{1/2}$};
\draw[level] (\levlength/2+0.2cm,-\ysep/3) -- ++(\levlengthd,0) node[midway,yshift=0cm,xshift=1.2cm] {$\mathrm{D}_{3/2}$};

\newdimen \bottomoffset \bottomoffset =.2cm
\newdimen \topoffset \topoffset =.5cm

\draw[virtual] (-\levlength/2,-\ysep+\bottomoffset) -- ++(\levlength,0) node[midway] (botmid) {};
\draw[virtual] (-\levlength/2,\ysep-\topoffset) -- ++(\levlength,0)  node[midway,yshift=1.5cm, xshift=-0.2cm] (topmid) {$^{\mathbf{138}}\mathbf{Ba}^\mathbf{+}$} ;

\draw[arrow] (-1.1,-\ysep) to[] ($(-1.1,\ysep*1.05)$) node[midway,yshift=0.4cm, xshift=-1.4cm] {$D_1$};
\draw[arrow] (\levlength/2+0.3cm,-\ysep/3) to[] ($(0.8,\ysep*1.05)$) node[midway,yshift=0.4cm, xshift=1.7cm] {$D_2$};

\draw[arrow] (0.2,-\ysep) to[bend left] ($(0.2,\ysep)+(0,-\topoffset)$) node[midway,yshift=-0.2cm, xshift=-0.6cm] {$\Omega_L$};
\draw[arrow] ($(0.2,\ysep)+(0,-\topoffset)$) to[bend left] ($(0.2,-\ysep*9/10)$) node[midway,yshift=-0.2cm, xshift=0.9cm] {$g$};

\coordinate (deltabottom) at ($(\levlength/2,-\ysep+\bottomoffset/2)$);
\coordinate (deltatop) at ($(\levlength/2,\ysep-\topoffset/2)$);

\node[xshift=0.5cm] at (deltabottom) {$\Delta_c$};
\node[xshift=0.4cm] at (deltatop) {$\Delta$};
\newdimen \dimheight \dimheight =.3cm
\draw[dimension] ($(deltabottom)+(0,\bottomoffset/2)+(0,\dimheight)$)--++(0,-\dimheight);
\draw[dimension] ($(deltabottom)-(0,\bottomoffset/2)-(0,\dimheight)$)--++(0,\dimheight);
\draw[dimension] ($(deltatop)+(0,\topoffset/2)+(0,\dimheight)$)--++(0,-\dimheight);
\draw[dimension] ($(deltatop)-(0,\topoffset/2)-(0,\dimheight)$)--++(0,\dimheight);

\end{tikzpicture}
\caption{\label{cavlevel} The relevant transitions and level structure for $^{138}\mathrm{Ba}^+$.  Doppler cooling is achieved by driving the $6\mathrm{S}_{1/2} \to 6\mathrm{P}_{1/2}$ transitions at $493\, \mathrm{nm}$ ($D_1$) and repumping on the $5\mathrm{D}_{3/2} \to 6\mathrm{P}_{1/2}$ transitions at $650\, \mathrm{nm}$ ($D_2$). The ion-cavity  coupling is driven by the cavity probing beam ($R_p$) with Rabi rate $\Omega_L$ and the intra-cavity field with coupling strength $g$. $\Delta$ is the detuning of the laser frequency from the $\mathrm{S}_{1/2} \leftrightarrow \mathrm{P}_{1/2}$ transition while $\Delta_c$ is the relative detuning between the laser and the cavity resonance. To obtain the ion-cavity emission profiles, $\Delta_c$ is swept $\pm 12 \, \mathrm{MHz}$ over the transition carrier ($\Delta_c=0$) while $\Delta$ is kept constant at $-110\,\mathrm{MHz}$. }
\end{figure}
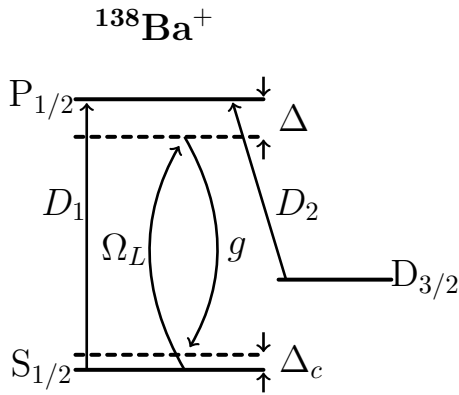

The experimental setup is illustrated in \Fref{iso} in which a high finesse cavity is aligned with its optical axis transverse to a linear Paul trap \cite{prestage,berkeland2,chuah}. Details of the ion trap have been reported elsewhere \cite{lewty}. Briefly, a $5.3 \, \mathrm{MHz}$ RF potential with an amplitude of $125 \, \mathrm{V}$ is applied via a step-up transformer to two diagonally opposing electrodes. A small DC voltage applied to the other two electrodes ensures a splitting of the transverse trapping frequencies and rotates the principle axes of the trap with respect to the propagation direction of the cooling lasers. Axial confinement is provided by two axial electrodes separated by $2.4 \, \mathrm{mm}$ and held at $33\, \mathrm{V}$. Using this configuration, we achieve trapping frequencies of $2 \pi \times (1.2, 1.1, 0.40) \, \mathrm{MHz}$ for a single $^{138}\mathrm{Ba}^+$ ion.

The relevant lasers and level structure for $^{138}\mathrm{Ba}^+$ are shown in \Fref{cavlevel}.  Doppler cooling is achieved by driving the $6\mathrm{S}_{1/2} \to 6\mathrm{P}_{1/2}$ transitions at $493\, \mathrm{nm}$ and repumping on the $5\mathrm{D}_{3/2} \to 6\mathrm{P}_{1/2}$ transitions at $650\, \mathrm{nm}$. The $493\, \mathrm{nm}$ cooling laser ($D_1$) and the $650\, \mathrm{nm}$ repumping laser ($D_2$) are both red-detuned by $\approx 15\, \mathrm{MHz}$ for optimum cooling. Both lasers are combined into a single fiber and sent into the trap along the $z$ direction defined by a 3 Gauss magnetic field.  The $D_1$ and $D_2$ beams are both linearly polarized perpendicular to the magnetic field to avoid unwanted dark states in the cooling cycle. The probe laser ($R_p$) is red-detuned by $110\,\mathrm{MHz}$ from the $\mathrm{S}_{1/2} \leftrightarrow \mathrm{P}_{1/2}$ transition and sent into the trap along the $x$ direction. $R_p$ is linearly polarized along the magnetic field direction and drives the cavity-induced Raman transition as illustrated in \Fref{cavlevel}.

The dual coated high finesse cavity is approximately $5\,\mathrm{mm}$ long with a finesse of $85000$ at $493\,\mathrm{nm}$ and $75000$ at $650\,\mathrm{nm}$. The cavity length is stabilized via Pound-Drever-Hall technique \cite{hall} to the sideband of a low linewidth $650 \, \mathrm{nm}$ laser \cite{chuah}. The sideband is generated by a wideband electro-optic modulator (EOM). Changing the EOM drive frequency allows us to tune the cavity resonance relative to the fixed frequency of the $650\,\mathrm{nm}$ locking laser. This laser is approximately $500\,\mathrm{GHz}$ detuned from the repump transition and thus does not impact on the cavity dynamics.  The probe laser ($R_p$) at $493\,\mathrm{nm}$ is referenced to the fixed frequency of the locking laser via a transfer cavity.  This ensures the probe laser has a well defined detuning relative to the cavity resonance. Cavity QED parameters relevant to the $493\,\mathrm{nm}$ probing transition are $(g,\kappa,\gamma)=2\pi\times(1.2,0.175,10.35)\,\mathrm{MHz}$ where $g$ is the cavity coupling strength for the $\mathrm{S}_{1/2}\leftrightarrow \mathrm{P}_{1/2}$  $\pi-$transition and $\gamma$ is the total dipole decay rate of the $\mathrm{P}_{1/2}$ level.

The cavity output is first passed through a dichroic mirror to separate the $493\,\mathrm{nm}$ output from the transmission of the $650\,\mathrm{nm}$ locking laser.  Further filtering is done using a bandpass filter with a specified transmission of $97\%$ at $493\,\mathrm{nm}$ and attenuation of $85\,\mathrm{dB}$ at $650\,\mathrm{nm}$.  The light is then coupled via a single mode fiber to a single photon counting module (SPCM).  From the transmission of the cavity at $493\,\mathrm{nm}$ ($24\%$), the fibre coupling efficiency ($70\%$), and the quantum efficiency of the SPCM at $493\,\mathrm{nm}$ ($45\%$) we estimate an overall detection efficiency of intra-cavity photons of approximately $7.5(2)\%$.

The cavity itself sits on an attocube nanopositioner which provides vertical adjustment of the cavity relative to the ion trap. Due to a small angular deviation of the cavity axis relative to the horizontal plane, this motion also results in a relative displacement of the ion along the cavity axis in an approximately $30:1$ ratio.  Thus a few micron vertical movement of the cavity allows us to move the ion from a cavity node to antinode without significantly altering the output coupling to the SPCM or the micromotion compensation.  In addition, the vertical displacement is much less than the mode waist ($\approx 40\,\mathrm{\mu m}$) and thus does not significantly alter the transverse alignment. By maximizing (minimizing) the scattering into the cavity we can locate the ion at the antinode (node) of the cavity to an accuracy of about $\pm 10\,\mathrm{nm}$ limited by the step size of the nanopositioner. To avoid heating of the ion during probing, we probe for just $200\,\mathrm{\mu s}$. With the ion maximally coupled to the cavity this results in the collection of $\sim 3-5$ photon counts near to the cavity resonance with a background of $\sim 0.1$ counts.  This is repeated 1000 times to give a total integration time of $0.2\,\mathrm{s}$. To ensure that the ion is equally cooled for every cycle of the measurement, $1\, \mathrm{ms}$ of Doppler cooling is used before each probing.

\section{Results} \label{sec4}

\begin{figure}
\centering
\includegraphics{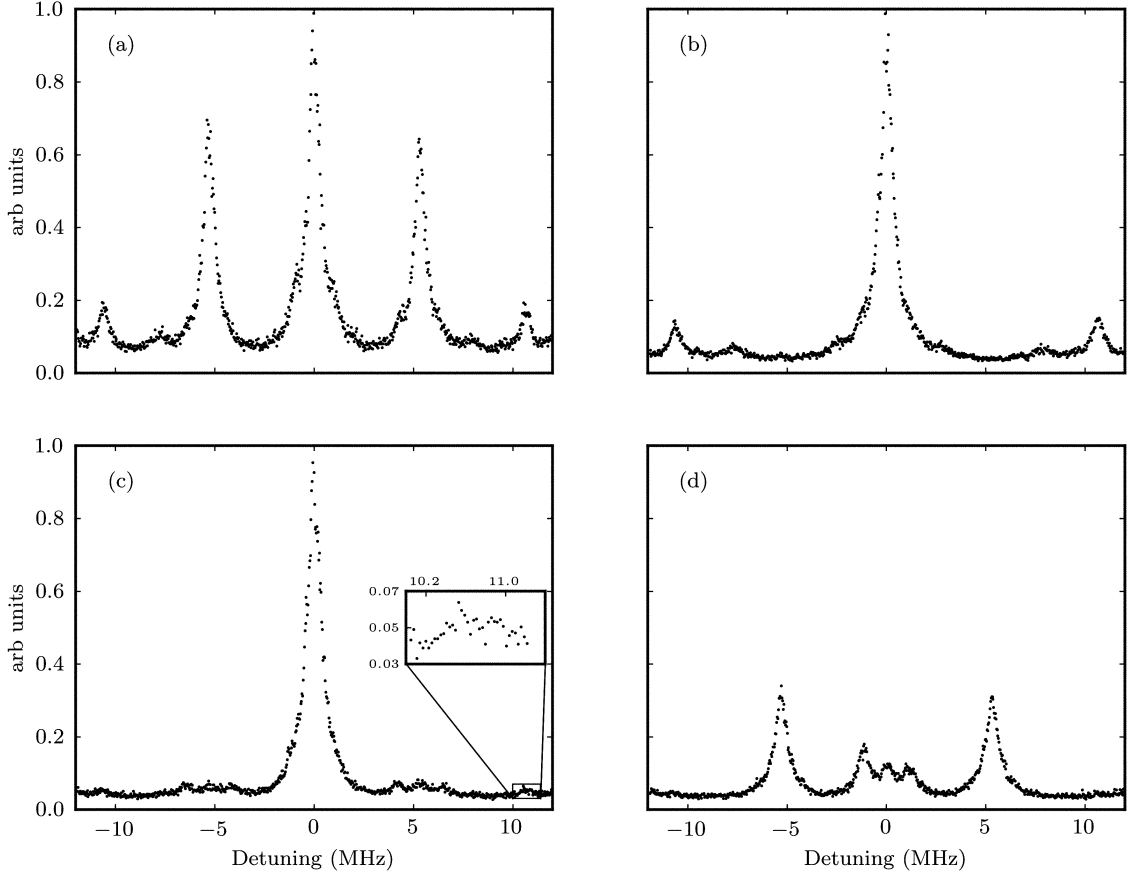}
 	 \caption{\label{micromotion} (a), (b) and (c) are the ion-cavity emission profiles obtained at the cavity anti-node while (d) is obtained at the cavity node. All plots are normalized to their respective carrier peaks except (d), which is normalized to the carrirer peak in (c). First and second order micromotion sidebands at $\pm\Omega$ and $\pm 2\Omega$ are clearly visible in (a) before any micromotion compensation. After compensating the micromotion along the probe direction ($\hat{\mathbf{x}}$), the first order sidebands are eliminated as shown in (b). The persistence of the second order sidebands at $\pm 2\Omega$ is due to the coupling of the micromotion along the cavity axis ($\hat{\mathbf{y}}$). Compensating the micromotion along this direction eliminates the second order peaks as shown in (c). For greater detection sensitivity, the ion is shifted to the cavity node. Consequently, the residual micromotion along the cavity axis manifests as sidebands with a much higher amplitude at $\pm\Omega$ as shown in (d). In the same plot, the peak at resonance is due to a residual offset from the cavity node.  The other two peaks are motional sidebands due to the secular motion of the ion.  }
\end{figure}

\begin{figure}
\centering
\includegraphics{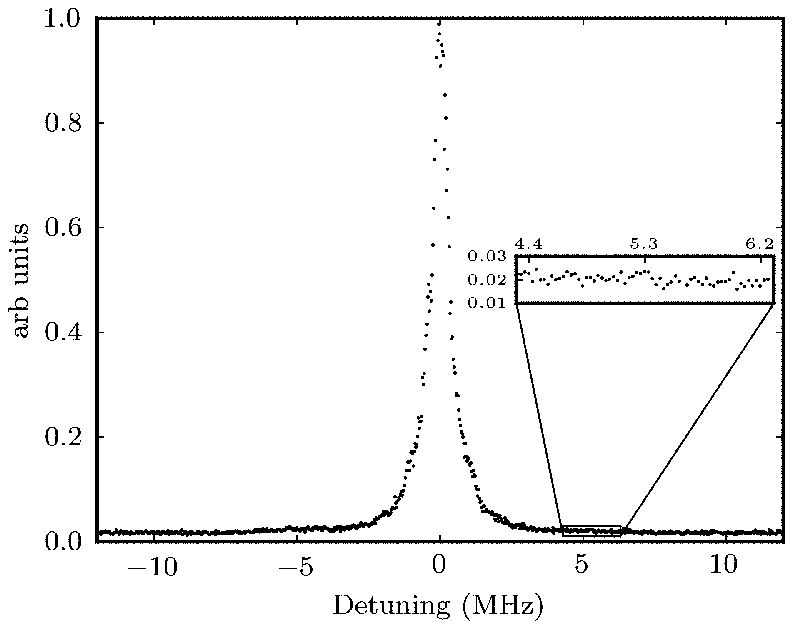}
 	 \caption{\label{micromotion2} The ion-cavity emission profiles for an ion located at the cavity anti-node with the micromotion fully compensated. The inset shows the data near to the RF sideband frequency which is statistically flat with no clear signature of a sideband present consistent with a signal to noise ratio of one.  }
\end{figure}

Typical emission spectrums are shown in \Fref{micromotion} and \Fref{micromotion2}. In \Fref{micromotion}(a) the ion is located at the antinode of the cavity.  First and second order micromotion sidebands at $\pm\Omega$ and $\pm 2\Omega$ are clearly visible. After compensating the micromotion along the probe direction the first order sidebands are eliminated as shown in \Fref{micromotion}(b). The second order sidebands still persist due to higher order terms that have been neglected in the expansion of \Eref{hamiltonian}. The neglected terms give rise to a sensitivity of the cavity emission to micromotion along the cavity axis. By compensating the micromotion along this axis these second order sidebands can also be elimnated as shown in \Fref{micromotion}(c). However, since this effect is higher order, use of the second sideband is much less sensitive to the micromotion amplitude along this direction and greater sensitivity is gained by shifting the ion to the node. This is evident by the spectrum in \Fref{micromotion}(d) taken after the ion is moved to the node. Residual micromotion along the cavity axis is still apparent from the presence of the first order sidebands allowing further micromotion compensation along that direction. We also note that the spectrum in this case contains three additional peaks near to resonance. The peak at resonance is due to a residual offset from the cavity node. The other two peaks are motional sidebands due to the secular motion of the ion. These peaks are unresolved in the previous figures due to the presence of the carrier.

With the micromotion fully compensated we obtain the spectrum shown in \Fref{micromotion2} which is taken with the ion located at the anti-node. The inset shows the data near to the RF sideband frequency which is statistically flat with no clear signature of a sideband present consistent with a signal to noise ratio of one. The data within the inset has a mean of $100$ counts with a standard deviation of $10$ and the maximum counts on the carrier is $5000$. Thus, from \Eref{betaxmin} we infer a minimum micromotion amplitude along the probe direction of $7.0(2)\,\mathrm{nm}$ which is approximately the spread of the ground state wave function along the transverse trap axes. Recently, micromotion compensation to the level of $1\,\mathrm{nm}$ has been reported in a $30\,\mathrm{s}$ integration time \cite{pyka}. Within the same integration time we would expect to improve our compensation to approximately $2.0\,\mathrm{nm}$.

With our present system there are a number of factors that limit the achievable compensation. Thermal motion of the ions reduces the effective cooperativity of the cavity \cite{leibrandt} in our case by a factor of $\sim 0.6$. This could be improved with better cooling or tighter confinement of the ion.  In addition, a small birefringence of the cavity exists which splits the cavity resonance for horizontal and vertical polarizations. Due to our limited optical access we can only probe at an angle of $45^\circ$ to the vertical. Thus the probe couples equally to both modes of the cavity. This reduces the effective scattering into the cavity also by a factor of $\sim 0.6$. Together, these two factors reduce the total signal by a factor of $2.8$ and hence the SNR by $1.7$. Finally, the two equally dominating factors that limit the background counts are dark counts from the SPCM ($\sim 250\,\mathrm{/s}$) and residual counts from the $650\,\mathrm{nm}$ locking beam ($\sim 250\,\mathrm{/s}$).  An additional filter would eliminate the counts from the locking beam and SPCMs with $15$ counts/s are available.  Thus our background could be reduced by a factor of $\sim 30$, improving the SNR by a factor of 2.4.  Altogether, a factor of 4 improvement in the micromotion compensation is therefore possible with our current system making sub-nanometer compensation possible. The micromotion along the $\hat{\mathbf{z}}$ axis has so far been neglected. Nonetheless, the micromotion minimization can be done easily by having an additional probing laser aligned to that axis.

In summary, we have presented a method to minimize excess ion micromotion which is well suited to cavity QED experiments equipped with high finesse cavities. Its applicability in the situation where only 2-D optical access is available also makes it a potentially useful technique in future micro-fabricated surface ion trap implemented with high finesse cavity \cite{chuang}.  We have also shown that sub nanometer micromotion compensation is readily achievable by this approach.  Such levels of compensation are important for precision metrology \cite{berkeland} and the study of atom-ion collisions \cite{doerk,nguy}, in which micromotion is a significant limiting factor.

We thank Markus Baden, Kyle Arnold and Andrew Bah for help with preparing the manuscript. This research was supported by the National Research Foundation and the Ministry of Education of Singapore.

\section*{References}

\bibliography{ref}

\end{document}